# Modeling novel oral nicotine use among adolescents

Mitchell C.[1*]; Barnett T.E.[1]; Newton M.[1]; Thompson E.L.[1], Livingston M.D.[2]

1. Department of Quantitative and Qualitative Health Sciences, School of Public Health, University of Texas at San Antonio
2. Behavioral Social and Health Education Sciences, Rollins School of Public Health, Emory University

## Abstract

Novel oral nicotine products, particularly nicotine pouches, have rapidly gained popularity among adolescents. Among U.S. high school students, nicotine pouch use has doubled since 2021, with 2.4% reporting current use in 2024. We analyzed Florida Youth Tobacco Survey data from 2022-2024 to assess prevalence trends and developed a grade-structured compartmental model to project future trajectories and evaluate intervention strategies. The model accurately captured observed trends across all high school grades and projected continued growth without intervention. We evaluated single and multi-parameter intervention strategies. Single-parameter interventions demonstrated limited effectiveness while multi-parameter strategies showed substantial effects. These findings underscore the need for comprehensive, multi-faceted interventions incorporating prevention education, cessation support, policy enforcement, and peer influence modification. Grade-specific targeting can enhance overall program effectiveness. School-based interventions should be implemented rapidly to address the accelerating epidemic of oral nicotine use among adolescents.

*: Corresponding Author: mitchellc4@uthscsa.edu"

**Introduction**

Novel oral nicotine products emerged on the United States (U.S.) market in 2016, mostly in the form of nicotine pouches, but also lozenges and gummies. Nicotine pouches contain either tobacco-derived or synthetic nicotine mixed with flavorings (e.g., mint, fruit) and other fillers (Robichaud et al., 2020). In the United States, 2.9% of adults reported using nicotine pouches and 0.4% reported current use of nicotine pouches in 2022 (Dai & Leventhal, 2024). Nicotine pouches are regulated as tobacco products and are by law are illegal for use by those under 21 (Tobacco to 21 Act, 2019); however, the 2024 National Youth Tobacco Survey found that among those under 18 years of age, 480,000 youth in the U.S. reported currently using nicotine pouches, which has doubled since 2021 (Park-Lee et al., 2024). In 2024, 4.7% of high school students reported ever using nicotine pouches, up from 3.1% in 2023 (Birdsey et al., 2023); 2.4% of high school students reported currently using nicotine pouches, up from 1.7% in 2023 (Jamal et al., 2024; Park-Lee et al., 2024). Of those currently using nicotine pouches, 29.3% reported frequent use; 22.4% of students reported daily use; and 86.5% reported using flavored pouches (Park-Lee et al., 2024). Available flavors include menthol, citrus, coffee, cinnamon, and fruit varieties, with the most sold flavor being mint (Marynak et al., 2021).

Results from a national online continuous tracking study of youth and young adults in the U.S. demonstrated that current nicotine pouch users were more likely to be 21 years or older, male, and lower income (Patel et al., 2023). Additionally, nicotine pouch use is reported to be higher among rural populations than urban populations for both adolescent and adult age groups (Han et al., 2025). Nicotine pouches have been marketed as being "tobacco-free," though they contain nicotine from tobacco leaves, and this might be interpreted by consumers as the "healthier" choice compared to traditional tobacco products (Van Thomme, 2025). However, just because these products are smokeless does not mean that they are harmless. Smokeless tobacco products like nicotine pouches can lead to nicotine addiction, diseases of the mouth and esophagus, tooth decay and tooth loss, reproductive and developmental risks when used during pregnancy, and risks of nicotine poisoning, heart disease, and stroke (Connolly et al., 2010; Fisher et al., 2005; Laldinsangi, 2022; Olivas et al., 2025; Piano et al., 2010).

Sales of nicotine pouches continue to rise each year regardless of the implications these products may have on health. Since nicotine products were first introduced into the global market in 2016, the U.S. has been a lead consumer. In August 2019, the U.S. reported 126 million nicotine pouches sold, and this number increased to 808 million by March 2022 (Majmundar et al., 2022). The nicotine pouch market was estimated to account for approximately $4.09 billion in 2024, according to Grand View Research (n.d.), and is projected to reach $33 billion by 2026 (Gupta & Mehrotra, 2021). Convenience stores account for over 97% of total sales (Marynak et al., 2021).

Given the recent increase in prevalence and sales, including the uptake among youth and young adults, we sought to forecast nicotine pouch use among a sample of high school students. In addition to modeling growth, we estimated potential intervention time points that could potentially slow or reverse the upward trend.

**Methods**

We assessed the prevalence of, and sociodemographic correlates to, novel oral tobacco use in a statewide youth tobacco surveillance system. We developed a grade-structured compartmental model to project epidemic dynamics and identify optimal intervention targets. The model was calibrated using three years of surveillance data (2022-2024) and then used to evaluate single and multi-parameter intervention strategies.

**Study Sample**

We used data from the Florida Youth Tobacco Survey (FYTS) (https://www.flhealthcharts.gov/ChartsDashboards/rdPage.aspx?rdReport=SurveyData.YTS.Dataviewer). The survey involves a 2-stage cluster probability sample design. First, a random sample of public middle schools and high schools (grades 6–12) is selected for participation in the survey. Second, within each selected school, a random sample of classrooms is selected, and students in those classrooms are invited to participate in the survey. Larger schools are sampled with greater probabilities of selection than smaller schools to ensure that every student in the state has the same probability of selection. Parental consent is typically obtained through passive permission forms that parents must return to opt-out their child from participation. In a few counties, however, active permission is required and is obtained (Barnett et al., 2009). FYTS uses a state-level representative sample in odd years and a county-level representative sample in even years. The current analysis focuses on high school students (grades 9-12) surveyed across three consecutive years: 2022 (school response rate 98.0%; student response rate 71.4%), 2023 (school response rate 77.1%; student response rate 56.7%), and 2024 (school response rate 86.0%; student response rate 68.7%). Sample sizes included substantial numbers of students across all demographic subgroups, providing adequate power to detect meaningful changes over time and differences between groups.

**Measures**

Sociodemographic variables used in this study included sex, grade, and race/ethnicity. The FYTS data report sex as a dichotomous variable, male or female. Race-ethnicity was established from two questions, 1) "How do you best describe yourself (Select only ONE response)?" with response options including American Indian or Alaska

Native, Asian, Black or African American, Native Hawaiian or Other Pacific Islander, White, or Other; and 2) "Are you Hispanic or Latino?" with response options Yes or No. FYTS combines these questions for one race/ethnicity variable that includes four categories: White, Black, Hispanic, Other. Grades were collected as response options including grades 6 – 12; only grades 9-12 (high school) were included in this study.

**Novel oral nicotine products**

The FYTS asked about "Other Nicotine Products" and provided definitions for each. The survey asked, "Have you ever tried, even once:" "Nicotine pouches?" or "Nicotine lozenges?" with yes and no as response options. Those who answered yes to either question were defined as ever users of oral nicotine products. The survey also asked, "During the past 30 days have you:" "Used nicotine pouches?" or "Used nicotine lozenges?" with yes or no as the two response options. Those who answered yes for either of these products were defined as current users of oral nicotine products.

**Background Data Analysis**

We assessed trends over three years among high school students from 2022-2024. The FYTS data reveal a consistent upward trajectory in oral nicotine product use. Ever use prevalence increased from 2.7% to 4.1%, while current use rose from 1.5% to 2.4% (Figure 1). Most striking is the universal nature of these increases—no demographic subgroup experienced decreases over the two-year period (Figure 2), indicating widespread adoption across the student population regardless of sex, grade level, or racial/ethnic background. This pattern suggests that novel oral nicotine products are rapidly becoming normalized among adolescents across all demographic segments.

**Methods: Model Development and Analysis**

To better understand and predict future trends in oral nicotine use among high school students, we developed a compartmental mathematical model that tracks the dynamic transitions of students between different behavioral states regarding nicotine use over time. This epidemiological modeling approach allows us to capture the complex social contagion dynamics of substance use behaviors and evaluate potential intervention strategies through mathematical simulation.

Our model employs a three-compartment structure for each grade level [citation for previous model], creating a total of twelve interconnected compartments across grades 9-12 (see Figure 1). Within each grade $i$, students are classified as potential users ($P_i$) who have never used oral nicotine products but remain susceptible to initiation, current nicotine users ($N_i$) who are actively using oral nicotine products, and quitters ($Q_i$) who previously used nicotine products but have ceased use. This grade-stratified approach recognizes that social influence patterns, cessation rates, and relapse behaviors may vary significantly across different developmental stages of adolescence (see Appendix for model details).

The mathematical framework incorporates several key behavioral assumptions. First, initiation into nicotine use occurs exclusively through social contagion mechanisms, where potential users transition to current users at a rate $\beta_i$ proportional to their contact with current users across all grade levels. This reflects the well-documented peer influence effects in adolescent substance use, where exposure to using peers increases individual risk of initiation (Marziali et al., 2022). The social influence parameter $\beta_i$ represents the effective contact rate between susceptible and using individuals, accounting for both direct social interactions and indirect influences through social media and peer networks.

Second, the model incorporates multiple pathways for cessation and relapse, reflecting the complex nature of nicotine dependence. Current users can quit through two distinct mechanisms: natural cessation at rate $\alpha_i$, representing individual decisions to quit independent of social influences, and socially influenced cessation at rate $\omega_i$, where contact with former users (quitters) encourages current users to quit. Finally, we model relapse behavior through dual pathways: natural relapse at rate $\lambda_i$, capturing the biological and psychological drivers of nicotine dependence that lead former users to resume use, and socially influenced relapse at rate $\delta_i$, where contact with current users triggers relapse among quitters.

The temporal dynamics of our system are governed by a system of ordinary differential equations that describe the rates of change in each compartment. For each grade $i$, the rate of change in potential users depends on the influx of new students entering that grade ($\mu$), the outflow due to initiation ($\beta_i P_i N$), and the natural progression to the next grade ($\mu P_i$). Current user dynamics incorporate initiation from potential users, relapse from quitters, cessation to the quitter compartment, and grade progression. The quitter compartment receives individuals from cessation processes while losing members to relapse and grade progression (See Table 1 for parameter descriptions and values).

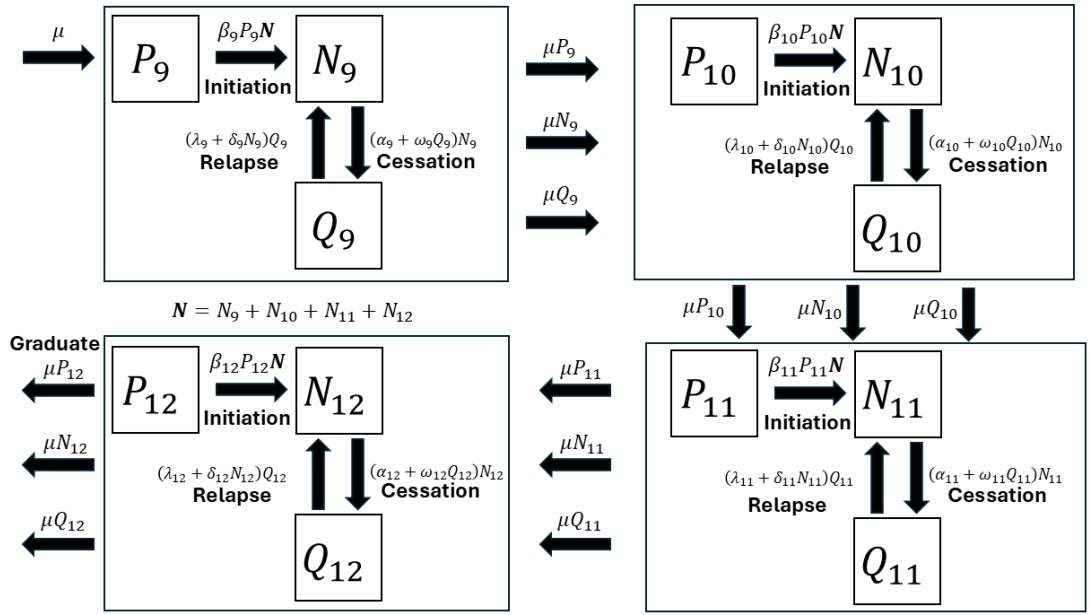

**Figure 1:** Simplified smoking behavior transition model across high school grades 9-12. The model shows potential smokers (P), current oral nicotine users (N), and quitters (Q) for each grade level, with transition rates between states and progression across grades. See methods for explanation of flows.

**Results**

Prevalence of ever use and current use increased across all demographic groups from 2022-2024. Year-over-year changes show widespread increases in 2022-2023 that accelerated in 2023-2024, with the largest recent increases among older students (Grade 11) and NH White populations. Black bars indicate increases; gray bars indicate decreases. (See Figure 2)

      Demographic differences emerged in both prevalence levels and rates of increase over time. Grade-related gradients were particularly pronounced, with older students consistently showing higher prevalence and larger increases. Students in grades 11-12 experienced the most dramatic two-year increases, with ever use increasing by more than 2.5 percentage points. This pattern suggests either age-related initiation occurring as students' progress through high school or cohort effects whereby certain student groups are more susceptible to adoption. Males demonstrated consistently higher usage rates and larger absolute increases than females across all time periods, with the male-female gap widening over time. Males showed 2.3 percentage point increases in ever use compared to 1.4 percentage points among females over the two-year period, indicating differential adoption patterns by sex. Substantial racial and ethnic differences were observed in both baseline prevalence and rates of change. Non-Hispanic (NH) White students exhibited the largest overall increases, with ever use rising by 2.9 percentage points over the two-year

period, while NH Black students showed the smallest changes at only 0.2 percentage points. Hispanic students demonstrated moderate increases of 0.7 percentage points, positioning them between the other two groups.

**Table 1: Parameters and descriptions for the oral nicotine model**, $i \in \{9, 10, 11, 12\}$

| Parameter | Description | Value | Source |
|---|---|---|---|
| $\mu$ | Progression through grades | 0.0027/day | Fixed |
| $\beta_i$ | rate of oral nicotine initiation | 0.1 - 0.2/day | fitted |
| $\lambda_i$ | rate of natural relapses in quitters | 0.01 - 0.8/day | fitted |
| $\delta_i$ | rate of socially influenced relapses in quitters | 0.4 - 0.9/day | fitted |
| $\alpha_i$ | rate of natural quitting for users | 0.3 - 0.9/day | fitted |
| $\omega_i$ | rate of socially influenced quitting for users | 0.2 - 0.6/day | fitted |
| $P_i$ | Potential oral nicotine users | | |
| $N_i$ | Current oral nicotine users | | |
| $Q_i$ | Oral nicotine quitters | | |
| $N$ | Total oral nicotine users | $N_9 + N_{10} + N_{11} + N_{12}$ | |

The year-over-year analysis reveals larger increases occurring in 2023-2024 compared to 2022-2023 for most demographic groups (Figure 2), suggesting the trend is gaining momentum rather than plateauing or decreasing. This acceleration is particularly pronounced among older students and NH White populations, indicating that the trend may be entering an exponential growth phase.

**FYTS Oral Nicotine Use: 2022-2024 High School Students**

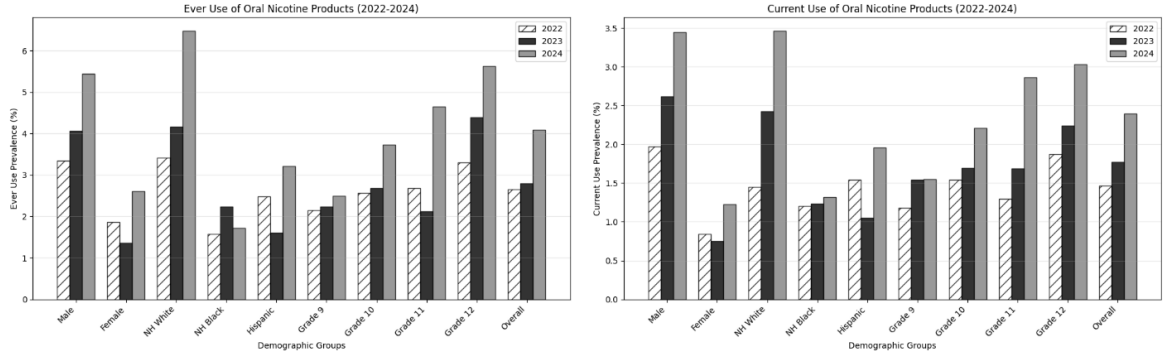

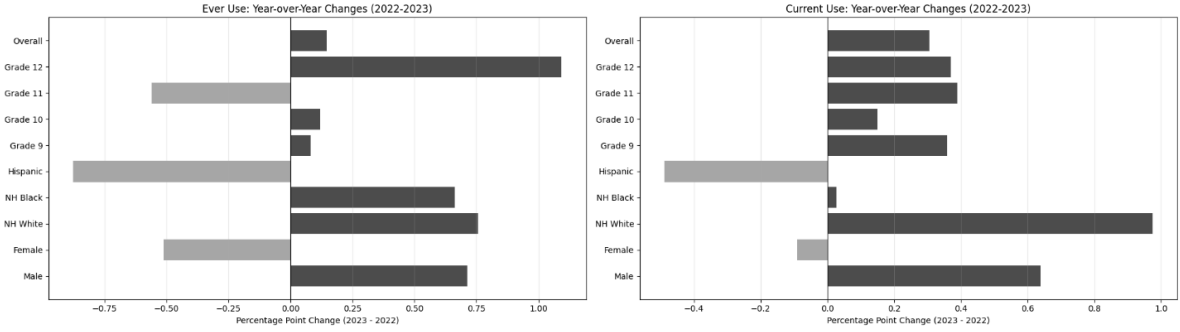

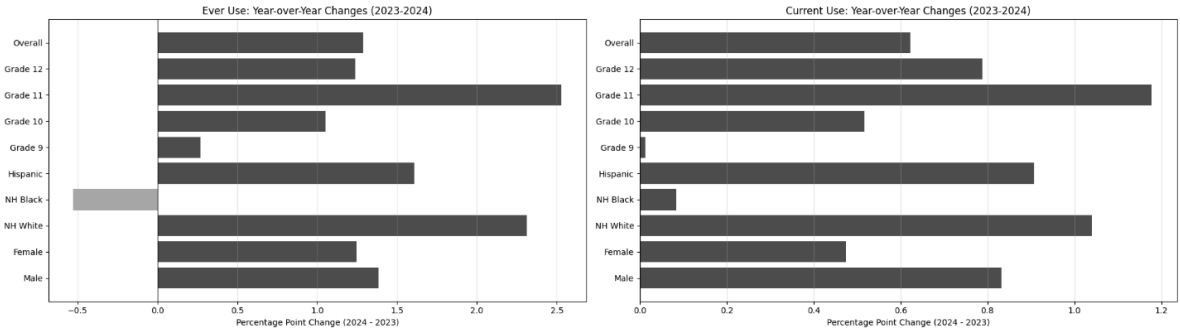

**Overall Changes: 2022-2024 (Two-Year Period)**

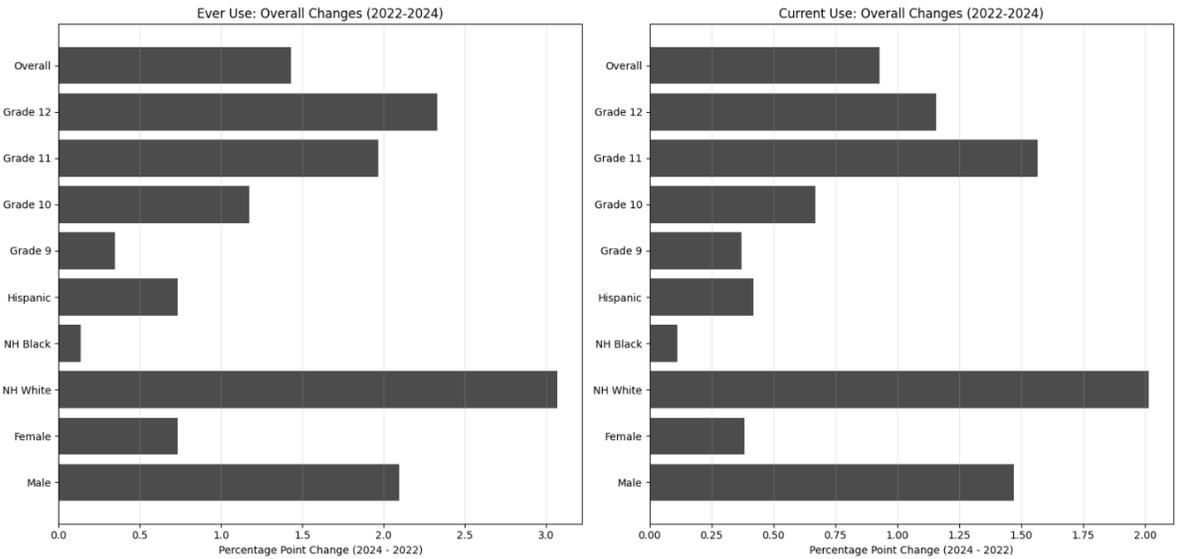

Figure 3: Percentage point increases in ever use (left) and current use (right) from 2022 to 2024, showing universal increases across all demographic groups. NH White students, older students (Grades 11, 12), and males experienced the largest two-year increases in both ever use and current use.

**Mathematical Framework and Stability Analysis**

We conducted a comprehensive stability analysis to determine the long-term behavior of our system under different parameter conditions. Our analysis focused on two equilibria that appear from the model: the Nicotine Free Equilibrium (NFE), representing the "disease-free" state where no students in any grade are currently using nicotine products, and the nicotine endemic equilibrium (NEE) which represents a more realistic representation with users. At the NFE equilibrium point, all potential users maintain their status ($P_i = 1$), while both current users and quitters are absent from all grades ($N_i = Q_i = 0$ for all $i$).

Using standard mathematical analysis techniques, we examined how our system behaves when it's close to the Nicotine Free Equilibrium to determine whether this state is stable or unstable. This analysis involves creating a simplified version of our complex system that's easier to analyze mathematically. To understand stability, we used mathematical tools that essentially test whether small disturbances to the nicotine-free state will grow larger (making the system unstable) or shrink away (keeping the system stable). Our analysis showed that some parts of the system always push toward stability - particularly the natural process of students graduating and leaving the high school population. However, other parts of the system, specifically how initiation and cessation interact within each grade level, determine whether nicotine use will persist or disappear. Each grade operates somewhat independently in terms of its stability properties, meaning we can analyze each grade separately to understand the overall system behavior.

The critical insight from our stability analysis is that each grade level exhibits independent stability properties with respect to the NFE, meaning that instability in any single grade can drive the entire system away from the nicotine-free state. The stability conditions require that the initiation rate $\beta_i$ must be sufficiently small relative to the sum of natural cessation ($\alpha_i$) and grade progression ($\mu$) rates. When $\beta_i$ exceeds a critical threshold, the NFE becomes unstable, indicating that nicotine use will persist and spread throughout the population.

Our analysis led to the derivation of the basic reproduction number for each grade $i$: $R_i = \beta_i(\lambda_i + \mu)/[\mu(\alpha_i + \lambda_i + \mu)]$. This quantity represents the expected number of secondary infections (new users) generated by a single current user during their entire period of use within grade $i$. The overall system reproduction number $R_0$ is defined as the maximum across all grade levels: $R_0 = max\{R_9, R_{10}, R_{11}, R_{12}\}$. When $R_0$ exceeds unity, the NFE becomes unstable, and the system will converge to an endemic equilibrium where nicotine use persists indefinitely in the population.

The reproduction number formula reveals important insights about intervention strategies. The numerator $\beta_i(\lambda_i + \mu)$ represents the total influence potential of a current user, incorporating both their ability to initiate new users and the average duration of their influence (determined by relapse rates and grade progression). The denominator $\mu(\alpha_i + \lambda_i + \mu)$ represents the rate at which users are removed from the system through cessation or grade progression. Effective interventions must either reduce initiation rates ($\beta_i$), increase cessation rates ($\alpha_i$), or reduce relapse rates ($\lambda_i$) sufficiently to bring $R_0$ below the critical threshold of one.

Notably, our stability analysis demonstrates that the social cessation and relapse parameters ($\omega_i$ and $\delta_i$) do not influence the stability of the NFE, though they significantly affect the dynamics and magnitude of endemic equilibria when they exist. This mathematical framework provides a rigorous foundation for understanding the threshold conditions under which nicotine use behaviors will persist or fade in high school populations and offers quantitative guidance for designing targeted intervention strategies.

**Model Validation and Intervention Performance**

Using Florida Youth Tobacco Survey data (2022-2024), weighted to represent the population of Florida students in public schools across the state, we validated the model by fitting it to grade-level and overall prevalence of ever and current users. With 2022 as initial conditions, the model accurately captured both grade-specific trends and overall epidemic trajectory through 2024. Model predictions showed excellent agreement with observed current use rates across all high school grades, with the baseline scenario projecting continued growth from 1.5% overall current use in 2022 to 5.5% by 2027 without intervention. Figure 4 shows the results. See Table 2 for best-fit values.

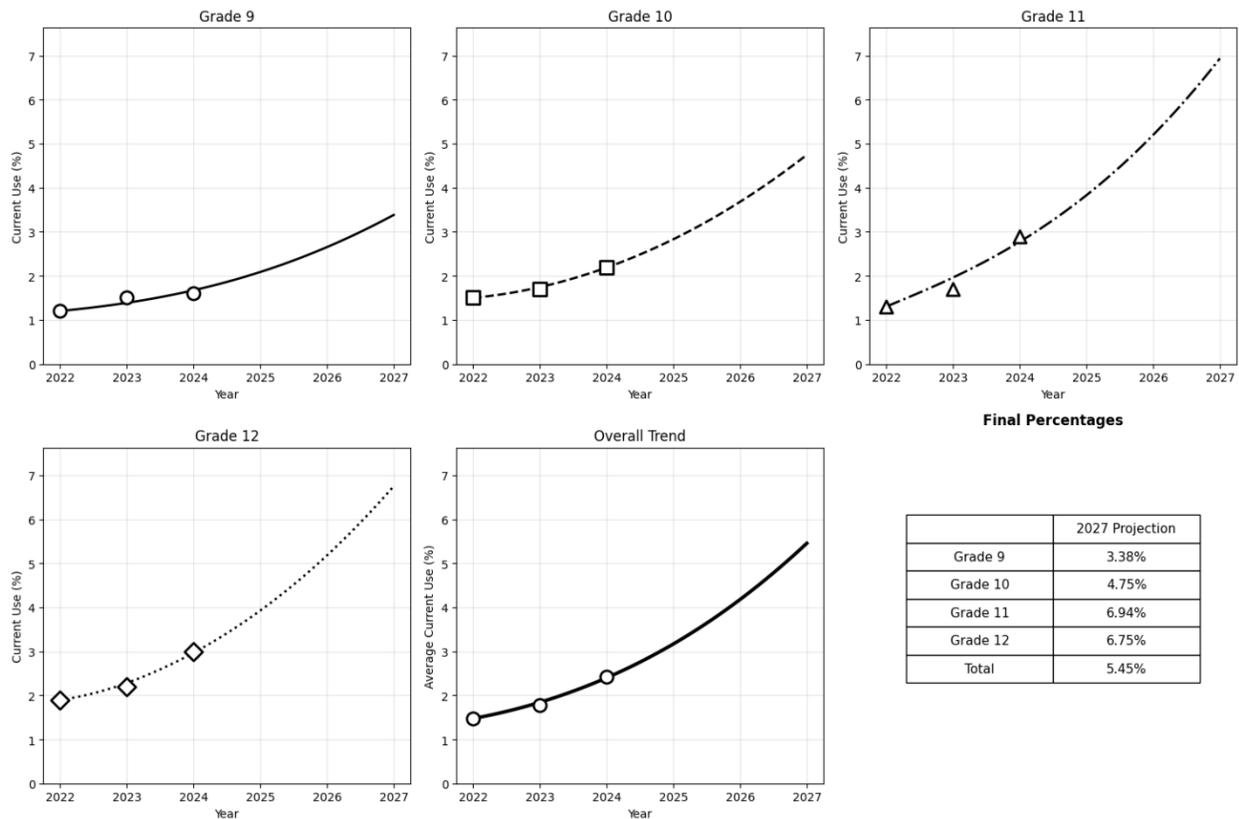

**Figure 4:** Model validation against Florida Youth Tobacco Survey data (2022-2024). Grade-specific trends showing model predictions and observed data for current tobacco use across high school grades 9-12. Overall epidemic trend comparing model predictions (solid line) with aggregated survey data (points). The model demonstrates excellent agreement with observed prevalence data across all grades and time points, with projections extending to 2027 under baseline conditions. Model fit captures both the grade-specific heterogeneity in oral nicotine user patterns and the overall increasing epidemic trajectory.

**Intervention Strategy Identification and Effectiveness**

We employed two complementary approaches to identify effective intervention strategies: random sampling of 100,000 parameter combinations with grade-specific targeting, and systematic evaluation of uniform multi-grade interventions. This dual approach reveals both optimized strategies for maximum effectiveness and practical school-wide programs for feasible implementation.

**Table 2: Best Fit Model Parameters from Florida Youth Tobacco Survey Data (2022-2024)**

| Parameter | Grade 9 | Grade 10 | Grade 11 | Grade 12 |
|---|---|---|---|---|
| $\beta_i$ (Transmission rate) | 0.1082 | 0.1881 | 0.2262 | 0.2486 |
| $\alpha_i$ (Cessation rate) | 0.4144 | 0.7491 | 0.8871 | 0.8950 |
| $\omega_i$ (Social cessation influence) | 0.8512 | 0.4584 | 0.0577 | 0.0105 |
| $\lambda_i$ (Relapse rate) | 0.0301 | 0.1537 | 0.4533 | 0.3073 |
| $\delta_i$ (Social relapse influence) | 0.3261 | 0.1801 | 0.7418 | 0.2938 |
| $\mu$ (Grade progression rate) | 0.0027 | 0.0027 | 0.0027 | 0.0027 |

**Optimized Grade-Specific Strategies**

Random parameter space exploration identified highly effective interventions achieving 85.7% to 94.7% total epidemic reduction using 5-8 parameters. The most effective strategy achieved 94.7% reduction through eight targeted modifications: initiation reduction in Grades 9-12 ($\beta_9$-80%, $\beta_{10}$-80%, $\beta_{11}$-80%, $\beta_{12}$-80%) and cessation enhancement in Grades 9, 10, 11, and 12 ($\alpha_9$+60%, $\alpha_{10}$+80%, $\alpha_{11}$+80%, $\alpha_{12}$+80%). This combination produced substantial grade-specific reductions with a 95% reduction for all grades.

We ran 100,000 samples through parameter space to determine the top parameters for achieving reduction in total usage. Across all combinations, 2-8, the only parameters that came up as the most effective were initiation and natural cessation. See Table 3 for results.

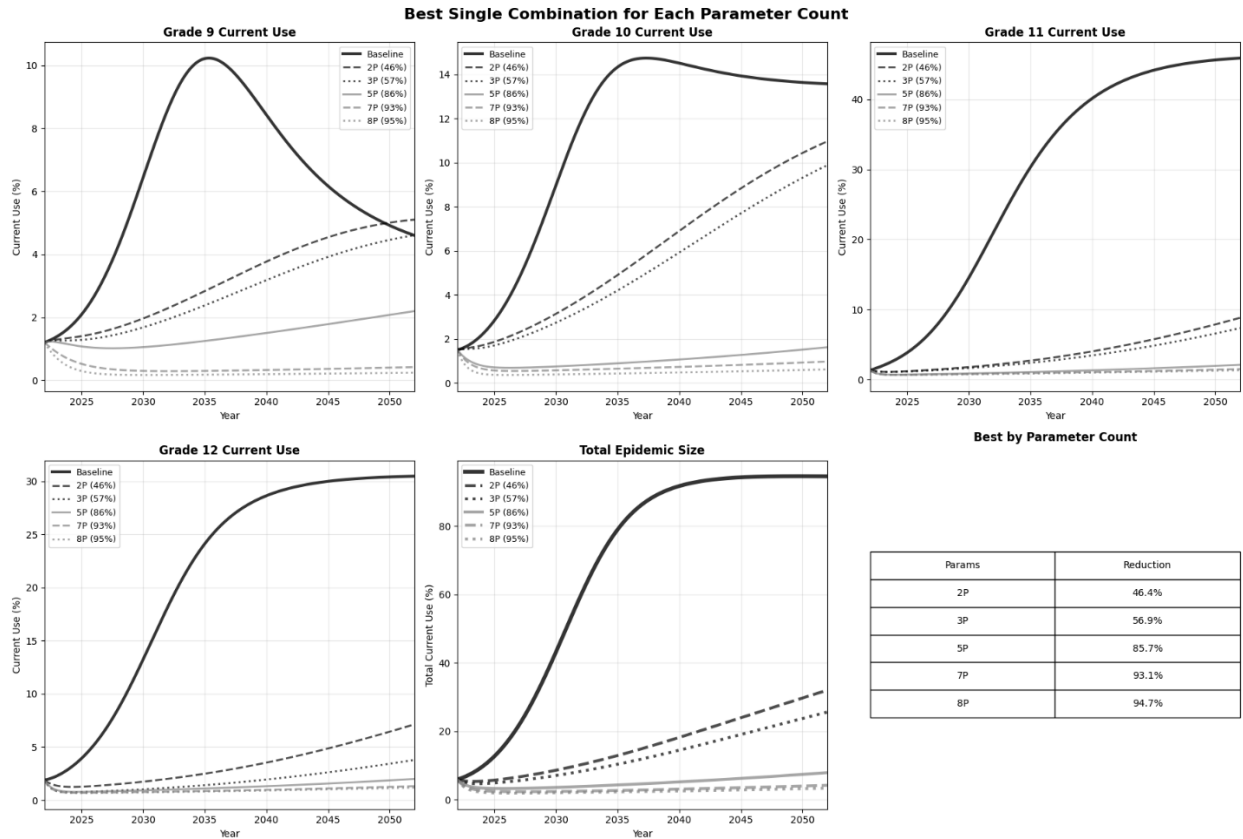

**Figure 5. Optimal Multi-Parameter Intervention Strategies by Complexity Level.** Grade-specific trajectories and total epidemic size under best-performing intervention combinations for each parameter count (2P through 8P). Higher parameter counts achieve progressively greater reductions, from 46.4% (2 parameters) to 94.7% (8 parameters). Higher grades (11-12) demonstrate greater responsiveness to interventions, with near-complete suppression under optimal 8-parameter strategies. Grade 9 shows more modest reductions, reflecting lower baseline transmission rates. The table (lower right) summarizes total epidemic reductions for each strategy complexity level.

Table 3: Parameter combinations appearing in intervention simulations.

| # of parameters used | Top parameter usage | Total percent reduction |
|---|---|---|
| 2 | β: 11; 12, α: 12 | 46.4% |
| 3 | β: 11, 12; α: 9, 10, 11, 12 | 56.9% |
| 5 | β, α: 9, 10, 11, 12 | 85.7% |
| 7 | β, α: 9, 10, 11, 12 | 93.1% |
| 8 | β, α: 9, 10, 11, 12 | 94.7% |

Single-parameter interventions demonstrated limited impact, with the most effective ($α_{11}$+20%) achieving only 6.1% reduction see Figure 6. However, multi-parameter combinations showed dramatic synergistic effects. Two-parameter strategies achieved 50.7% reduction, five-parameter strategies achieved 78.6% reduction, and the optimal eight-parameter strategy achieved 94.3% reduction. For practical planning, achieving 30% total reduction required only 2 parameters, while the ambitious 70% reduction target required at least 4 parameters.

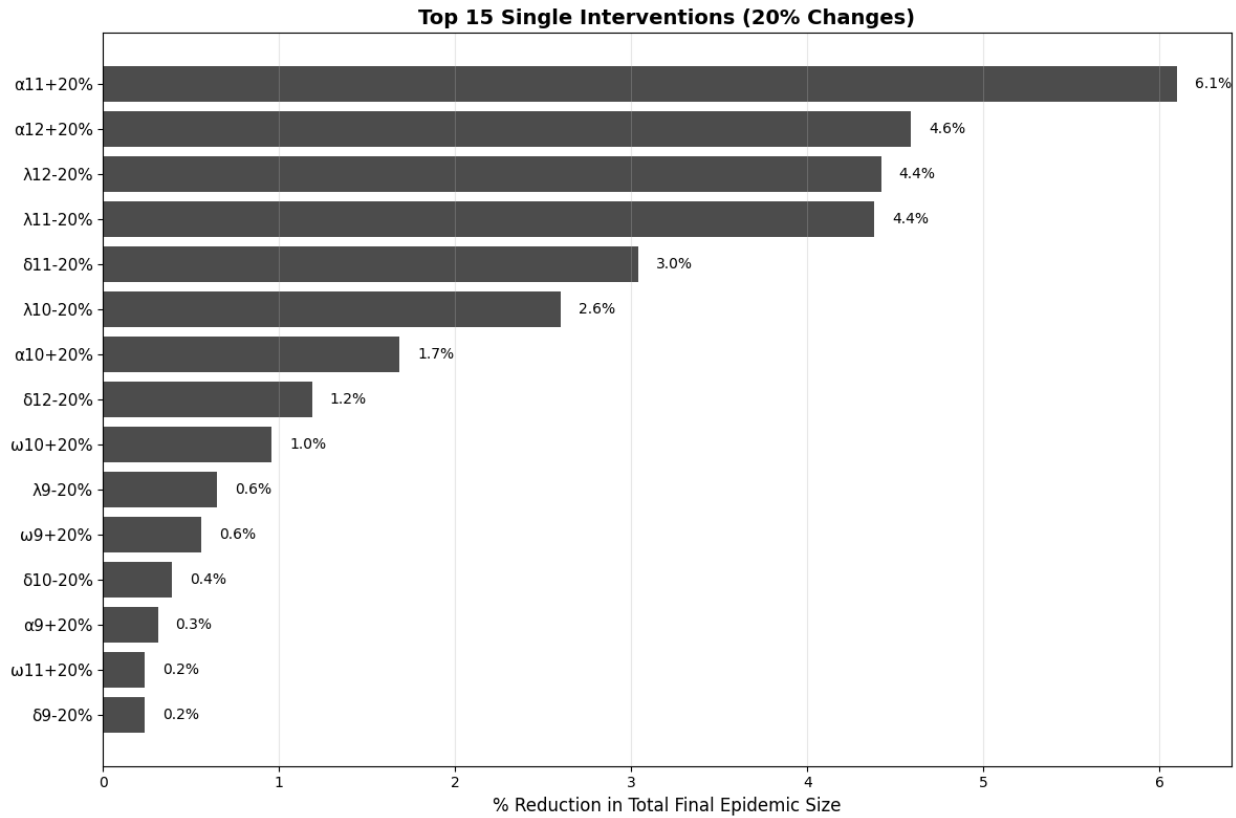

**Figure 6: Effectiveness of Single-Parameter Interventions on Total Epidemic Size.** Individual 20% parameter modifications demonstrate limited impact on overall epidemic control, with the most effective intervention (Grade 11 cessation enhancement, $α_{11}$+20%) achieving only 6.1% total reduction. Higher grade interventions targeting cessation (α) and initiation (λ) parameters show greater effectiveness than lower grade or alternative parameter modifications. The modest effects of single interventions underscore the necessity of multi-parameter strategies for substantial epidemic control.

Intervention effectiveness varied substantially by grade level, consistent with earlier analysis. Higher grades (11-12) demonstrated greater responsiveness to transmission reduction interventions, with several strategies achieving >90% reductions. Grade 9 showed modest responses to transmission reduction but greater sensitivity to cessation enhancement, with top interventions achieving 25-51% reductions. Grade 10 exhibited intermediate responsiveness (26-86% reductions). These patterns reflect grade-specific $R_0$ values, where higher grades contribute disproportionately to overall epidemic dynamics through elevated transmission rates.

**Uniform School-Wide Strategies**

Uniform interventions applying identical parameter changes across all grades offer simpler implementation while maintaining strong effectiveness. At 50% intervention intensity, five-parameter combinations (β, α, ω, λ, δ) achieved 88.7% total epidemic reduction—approaching optimized strategies but with uniform programming. Moderate 30% interventions produced 48.0% reduction with five parameters or 35.4% reduction with two parameters (cessation and initiation). Even modest 10% intervention levels achieved 12.4-15.5% reductions depending on parameter coverage. See Figure 7.

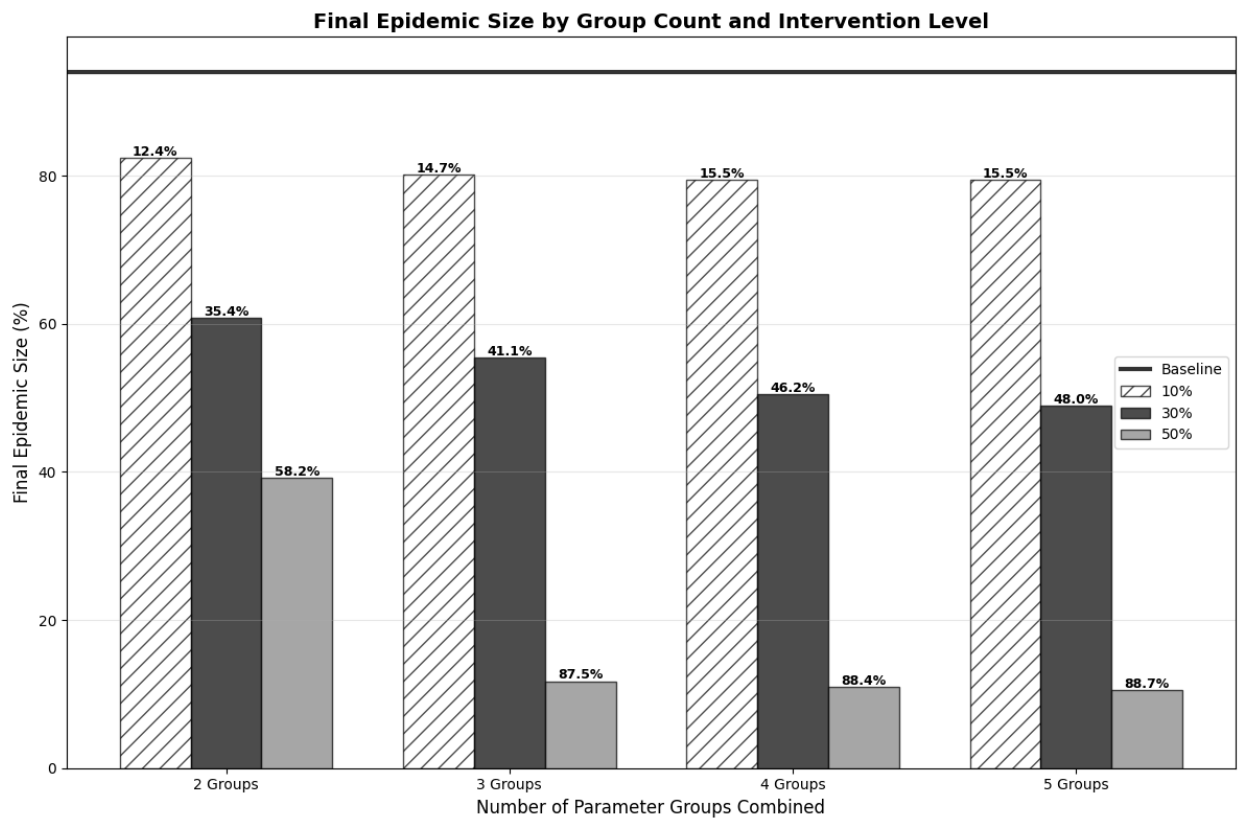

**Figure 7. Effectiveness of Uniform Multi-Grade Interventions by Coverage and Intensity.** Total epidemic size under coordinated school-wide strategies that apply identical parameter changes across all grades simultaneously. Percentages on bars are the total reduction from baseline. Effectiveness increases with both the number of parameter

groups combined (2-5 groups) and intervention intensity (10%, 30%, 50%). Five-parameter combinations at 50% intensity achieve 88.7% reduction, approaching the effectiveness of optimized grade-specific strategies while offering simpler implementation. Even moderate 30% interventions with comprehensive five-parameter coverage achieve 48.0% reduction, demonstrating the viability of practical school-wide programming.

**Implementation Examples**

These mathematical interventions may translate into real-world, school-based programs. Transmission reduction ($\beta$) corresponds to tobacco-free campus policies, reduced product visibility, and peer influence modification through social norms campaigns. Cessation enhancement ($\alpha$) represents school-based quit programs and counseling services. Initiation reduction ($\lambda$) includes prevention education, retail restrictions near schools, and flavor bans targeting youth appeal. Social influence modification ($\omega$) involves peer-led interventions and social marketing campaigns. Relapse prevention ($\delta$) encompasses ongoing support groups and quit maintenance programs.

**Discussion**

Like other studies (Jamal et al., 2024; Park-Lee et al., 2024; Birdsey et al., 2023), we found that novel oral nicotine use is increasing among adolescents. The consistent pattern of larger recent increases across multiple demographic groups suggests that environmental factors—such as increased marketing, product availability, or social media influence—may be driving accelerated adoption. These temporal patterns point to the emergence of high-risk groups among older, male, and NH White students who may require immediate targeted intervention strategies to prevent further escalation of use.

The results of the modeling indicate that without intervention, the prevalence and rapid growth of these products will continue. The model, however, provides projections for modest intervention and the resulting potential decrease in use. For example, a realistic 30% five-parameter school-wide strategy might include: strengthening and enforcing tobacco-free policies, implementing grade-appropriate cessation programs, enhancing prevention curricula, launching peer-led social norms campaigns, and establishing ongoing support systems. This coordinated approach could achieve 48.0% epidemic reduction with uniform implementation across all grades, making it administratively feasible while maintaining substantial effectiveness.

Given the newness of these products, we have only three years to validate the mathematical model. However, researchers found that oral nicotine pouch sales are rising exponentially, with monthly unit sales increasing from 327 million in July 2021 to 1,046 million in May 2024 in the U.S. He et al. (2025). Electronic cigarettes demonstrated a similar rapid increase from 2016 to 2019, with current use increasing from 11.3% to 27.5%.

Youth and young adults under 21 are not allowed to be directly marketed to for nicotine pouch advertising but may still encounter the advertising. In 2019, the top six brands of nicotine pouches spent a total of $11.2 million on advertising (Ling et al., 2023). Duan and colleagues (2024) found that the greatest proportion of ad occurrences and expenditures for nicotine pouches were television and radio, followed by online and mobile. The brand Zyn has previously been linked to social media influencers promoting their nicotine pouches, although the brand states that they have not paid these influencers for their promotions (Clark, n.d.). These influencers have been dubbed "Zyn-fluencers" and include well-known personalities such as Tucker Carlson, Joe Rogan, and Emma Chamberlain, for example. Zyn-fluencers and online media describe how the nicotine in pouches produces cognitive benefits, can lead to weight loss, and can be used as a smoking cessation product, though nicotine pouches have not been approved by the U.S. Food and Drug Administration for any of these uses (Bomey, 2025; Centers for Disease Control and Prevention, 2025; Nebraska Medicine, 2025).

One large driver for increasing sales is how brands like Zyn have been gamified. Zyn users can gain points in their Zyn Rewards account by purchasing a 15-pouch cannister, scanning a barcode on the bottom of the cannister, and cashing in these points to claim prizes (Dobbs et al., 2024). Participants have the opportunity to claim gift cards, electronics, outdoor gear, home appliances, and more using their rewards points. Public health experts and regulatory agencies should consider banning gamification of all tobacco and nicotine products given such practices encourage overconsumption to earn points. The practice of gaming often targets vulnerable populations, like young adults and adolescents, who can be attracted to the idea of a rewards system and then are introduced to frequent and normalized nicotine use without understanding how it may affect their health (Demopoulos, 2024; Truth Initiative, 2024).

**Limitations**

The study is not without limitations. We limited the analysis to one state surveillance system. The data, however, are publicly available and timely, Florida provides data on the most recent tobacco products in a timely manner. We also only had three years to fit the model, based on the relative new uptake of these novel products. We did not model by all demographic variables as this would introduce too much heterogeneity into the model. The results emphasizing grade differences likely represent an opportunity for focused interventions, whereas programs targeting by sex or race for intervention may not be feasible.

**Conclusion**

There is a clear need for prevention and cessation interventions, incorporating the role of social/peer influence. Although this is aligned with other substance use in adolescent populations and many of the same interventions should work, they need to be implemented faster when new products emerge. While most schools would and should continue to implement a school-wide intervention, some of the modeling identified distinct grade levels that might benefit from different types of interventions (e.g., 12$^{th}$ grade social influence for transmission reduction and 9$^{th}$ grade a focus on cessation). This might reflect developmental time periods, whereas older students are using in social situations more, and younger students may have higher levels of addiction and represent higher use youth or risk takers. Interventions should be multifaceted, and targeted additional interventions could strengthen the success of an overall program.

Appendix:

**Model Development**

To utilize this data and forecast future trends a model was created for oral nicotine use among high school students in Florida. The model partitions the population into three compartments: potential users ($P$), current nicotine users ($N$), and quitters ($Q$), using previous modeling framework by Machado-Marques and Moyles. These are extended further by considering high school grades for each compartment group, 9-12th grade. The model assumptions are carried over from previous work with novel products: Initiation of novel oral nicotine behaviors predominantly occurs among adolescents from social contagion from current novel oral nicotine users across all grades. All individuals age into the 9th grade as potential oral nicotine users but progress from their individual groups to the next grade, i.e. $P_9$ progresses to $P_{10}$ and so on. The size of the high school population remains constant over time, meaning high school students age into and out of the population at the same rate. The high school population is well-mixed, although this mixing does not strictly require close physical proximity. For example, it is possible to make social contact with others and potentially influence them over the internet. i.e. $P_9$ progresses to $P_{10}$ and so on. The size of the high school population remains constant over time, meaning high school students age into and out of the population at the same rate. The high school population is well-mixed, although this mixing does not strictly require close physical proximity. For example, it is possible to make social contact with others and potentially influence them over the internet.

Adolescents who are potential oral nicotine smokers ($P_i, i \in \{9,10,11,12\}$) only begin using it due to social contagion from oral nicotine user peers ($N_i$) at a rate $\beta_i > 0$. Once a high school student is using, they quit of their own volition at a rate $\alpha_i > 0$ or quit due to social influences from quitters ($Q_i$) at a rate $\omega_i > 0$. Those who become quitters' relapses and become oral nicotine users again from nicotine dependency and other non-social/natural factors at a rate $\lambda_i > 0$, or due to social contagion from other oral nicotine users at a rate $\delta_i > 0$. Further, individuals age into and out of grades at a rate $\mu_i > 0$. As was done in Alkhudari et al., a normalized population size is used such that $P_i + N_i + Q_i = 1$, where $P_i, N_i,$ and $Q_i \geq 0$ represent proportions of the adolescent population. See Table 1 for parameters. The equations for this model are: $Q_i \geq 0$ represent proportions of the adolescent population. See Table 1 for parameters. The equations for this model are:

**Mathematical Model Equations:**

$$dP_i/dt = \mu_i - \beta_i P_i N - \mu_i P_i$$

$$dN_i/dt = \beta_i P_i N + (\lambda_i + \delta_i N_i)Q_i - (\alpha_i + \omega_i Q_i + \mu_i)N_i$$

$$dQ_i/dt = (\alpha_i + \omega_i Q_i)N_i - (\lambda_i + \delta_i N_i + \mu_i)Q_i$$

$$N = N_9 + N_{10} + N_{11} + N_{12}$$

Each movement out at the rate $\mu_i$ progresses to the next grade while the 12th graders graduate and leave the population. See Figure 3 for the flow diagram and Table 1 for parameters.

## Mathematical Framework and Stability Analysis

The model has a Nicotine Free Equilibrium ($NFE$), $(P_i, N_i, Q_i) = (1,0,0)$ for each grade $i \in \{9,10,11,12\}$. Using standard Jacobian analysis, we linearized around the NFE and evaluated eigenvalues to determine stability conditions. Four eigenvalues equal $-\mu$ (always negative), while the remaining four are given by the reduced matrix:

$$\begin{bmatrix} \beta_i - (\alpha_i + \mu) & \lambda_i \\ \alpha_i & -(\lambda_i + \mu) \end{bmatrix}$$

where $i \in \{9,10,11,12\}$. Each grade has independent stability conditions with no cross-grade interactions affecting NFE stability. Notably, social cessation and relapse parameters ($\delta_i$ and $\omega_i$) do not influence stability.

For stability, we require negative trace and positive determinant. When $\beta_i < \mu$, the NFE is stable, indicating initiation rates below grade advancement rates ensure students age out faster than tobacco influence spreads. Otherwise, we define the basic reproduction number:

$$R_i = \frac{\beta_i(\lambda_i + \mu)}{\mu(\alpha_i + \lambda_i + \mu)}$$

with $R_0 = max\{R_9, R_{10}, R_{11}, R_{12}\}$. When $R_0 > 1$, the NFE becomes unstable, requiring intervention across all grades. $R_i$ represents secondary users one current user from grade $i$ influences during their usage period.

Sampling of 20,000 intervention combinations identified highly effective strategies targeting key model parameters. The top 20 intervention strategies achieved total epidemic reductions ranging from 73.3% to 91.5%, with parameter requirements between 6-8 interventions. Consistent with theoretical analysis, the most effective interventions focused on reducing initiation rates ($\beta_i$) and increasing cessation rates ($\alpha_i$) across grades.

The most effective strategy achieved 91.5% total reduction through 8-parameter modification:

$$\delta_{10}: -80\%, \ \alpha_{10}: +20\%, \ \lambda_{10}: -80\%, \ \beta_{12}: -80\%, \ \lambda^{11}: -80\%, \ \lambda^{12}: -80\%, \ \beta^{11}: -80\%, \ \alpha_9: +80\%$$

. This combination produced substantial grade-specific reductions: Grade 9 (51.2%), Grade 10 (82.6%), Grade 11 (95.2%), and Grade 12 (94.3%), demonstrating the effectiveness of targeting both transmission reduction ($\beta_i$) and cessation enhancement ($\alpha_i$) parameters identified in the stability analysis.

For practical implementation guidance, we identified minimum intervention requirements for specific reduction targets. Achieving >30% total epidemic reduction required only 2 parameters with 50.5% effectiveness, while the ambitious 70% reduction target required 5 parameters with 72.6% effectiveness. Intervention efficiency varied from 9.2% to 13.1% reduction per parameter, with the most efficient strategy achieving 13.1% reduction per parameter using only 6 interventions.

Theoretical analysis suggests two distinct intervention approaches: when grade-specific initiation rates satisfy $\beta_i < \mu$, focusing solely on initiation reduction can achieve stability as students age out faster than tobacco influence spreads. However, when $R_0 > 1$, comprehensive multi-grade interventions targeting both initiation ($\beta_i$ reduction) and cessation ($\alpha_i$ enhancement) become necessary for epidemic control.

Intervention effectiveness varied substantially across grade levels, with patterns consistent with reproduction number analysis. Higher grades (11-12) consistently demonstrated greater responsiveness to $\beta_i$ reduction interventions, with several strategies achieving >90% reductions in these populations. Grade 9 showed more modest responses to transmission reduction but greater sensitivity to cessation enhancement ($\alpha_9$ increases), with the most effective interventions achieving 25-51% reductions. Grade 10 exhibited intermediate responsiveness, with top strategies achieving 26-86% reductions. These patterns reflect the grade-specific $R_i$ values, where higher grades contribute disproportionately to overall $R_0$ through elevated transmission rates, making them priority targets for initiation-focused interventions.

Analysis revealed that multi-parameter interventions targeting transmission reduction ($\beta_i$), cessation enhancement ($\alpha_i$), and relapse prevention ($\lambda_i$) provided synergistic effects exceeding individual parameter impacts. C consistent with stability analysis showing $R_0$ reduction through lowering $\beta_i$ or increasing $\alpha_i$, the most successful strategies prioritized these parameters. The most effective interventions combined grade-specific parameter modifications rather than uniform across-grade approaches, with particular emphasis on reducing initiation rates in higher grades ($\beta_{11}, \beta_{12}$) and enhancing cessation in Grade 9 ($\alpha_9$). Notably, social parameters ($\delta_i, \omega_i$) played supporting roles but were not primary drivers of epidemic reduction, aligning with theoretical findings that these parameters do not affect $NFE$ stability.